\documentclass[twocolumn]{autart}    
\usepackage[british]{babel}
\usepackage{amsmath}
\usepackage{comment}
\usepackage{color}
\usepackage{amsfonts}
\usepackage{graphicx}          
\begin{document}

\begin{frontmatter}

\title{Decentralized Periodic Event-Triggered Control with Quantization and Asynchronous Communication\thanksref{footnoteinfo}} 

\thanks[footnoteinfo]{This work is partly funded by China Scholarship Council (CSC). Corresponding author A.~Fu. Tel. +31-015-27-83371.}

\author{Anqi Fu}\ead{A.Fu-1@tudelft.nl},    
\author{Manuel Mazo, Jr.}\ead{M.Mazo@tudelft.nl}  


\address{Delft Center for Systems and Control, Delft University of Technology, The Netherlands}


\begin{abstract}                          
Asynchronous decentralized event-triggered control (ADETC) \cite{MazoJr2014} is an implementation of controllers characterized by decentralized event generation, asynchronous sampling updates, and dynamic quantization. Combining those elements in ADETC results in a parsimonious transmission of information which makes it suitable for wireless networked implementations. We extend the previous work on ADETC by introducing periodic sampling, denoting our proposal asynchronous decentralized periodic event-triggered control (ADPETC), and study the stability and $\mathcal{L}_2$-gain of ADPETC for implementations affected by disturbances. In ADPETC, at each sampling time, quantized measurements from those sensors that triggered a local event are transmitted to a dynamic controller that computes control actions; the quantized control actions are then transmitted to the corresponding actuators only if certain events are also triggered for the corresponding actuator. The developed theory is demonstrated and illustrated via a numerical example.
\end{abstract}
\end{frontmatter}

\section{Introduction}
\label{section:introduction}

In digital control applications, the control task consists of sampling and transmitting the output of the plant, computing and implementing controller outputs. Current developments of sensor and networking technologies have enabled the emergence of wireless networked control systems (WNCS), in which communication of distributed components is established via wireless networks. WNCS can be established and updated with large flexibility and low cost, and are especially suitable to physically distributed plants. Limited energy supplies are often the case when sensors are battery powered for mobility and/or flexibility reasons. The major challenge in WNCS design is thus to achieve prescribed performance under limited bandwidth and energy supplies. Our present work is mostly inspired by \cite{heemels2013periodic}, \cite{liberzon2007input}, and \cite{MazoJr2014}. In \cite{heemels2013periodic}, Heemels et. al. present a periodic event-triggered control (PETC) mechanism. In PETC, the sensors sample the output of the plant and verify the central or local event conditions periodically. Therefore, the energy consumed by sensing is reduced compared to those continuously monitoring event-triggered mechanisms, while still a pre-designed performance can be guaranteed. In \cite{liberzon2007input}, Liberzon and Ne\v{s}i\'{c} present a state dependent quantizer which zooms in and out based on the system's state, so as to provide input to state stability (ISS).
In \cite{MazoJr2014}, Mazo and Cao present an asynchronous decentralized event-triggered control (ADETC) mechanism combining state dependent dynamic quantization and decentralized event-triggering conditions.
We propose an asynchronous decentralized periodic event-triggered control (ADPETC) mechanism building on the aforementioned pieces of work with the goal of reducing wireless channel bandwidth occupation and energy consumption. This ADPETC incorporates: quantization in a zooming fashion, which is similar to \cite{liberzon2007input} and \cite{MazoJr2014}; an asynchronous event-triggered mechanism, based on \cite{MazoJr2014}; and periodic sampling as in \cite{heemels2013periodic}. Moreover, compared with \cite{liberzon2007input} and \cite{MazoJr2014}, in our approach the quantization error or global threshold depends on the information in the controller, instead of just on the current estimation of the system's state; compared with \cite{heemels2013periodic}, in which the algorithm for designing decentralized event condition parameters is complex: requiring to solve a set of linear matrix inequalities (LMIs), our approach requires to solve only one LMI. Compared with our preliminary version \cite{fu2016periodic}, the main difference is, in \cite{fu2016periodic}, a set of bilinear matrix inequalities (BMIs) are required to solve when design the event condition parameters. In our current work, the event condition can be less conservative, thus resulting in less triggered events.

\section{Preliminaries and problem definition}

We denote the positive real numbers by $\mathbb{R}^+$, by $\mathbb{R}^+_0=\mathbb{R}^+\cup\{0\}$, and the natural numbers including zero by $\mathbb{N}$. $|\cdot|$ denotes the Euclidean norm in the appropriate vector space, when applied to a matrix $|\cdot|$ denotes the $l_2$ induced matrix norm. 

Let us consider a linear time-invariant (LTI) plant given by:
\begin{equation}\label{eq:system}
\left\{
\begin{aligned}
  \dot{\xi}_p(t)&=A_p\xi_p(t)+B_p\hat{v}(t)+Ew(t)\\
  y(t)&=C_p\xi_p(t),\\
\end{aligned}
\right.
\end{equation}
where $\xi_p(t)\in\mathbb{R}^{n_p}$ and $y(t)\in\mathbb{R}^{n_y}$ denote the state vector and output vector of the plant respectively, $\hat{v}(t)\in\mathbb{R}^{n_v}$ denotes the input applied to the plant, $w(t)\in\mathbb{R}^{n_w}$ denotes an unknown disturbance. The plant is controlled by a discrete-time controller given by:
\begin{equation}\label{eq:controllaw}
\left\{\begin{aligned}
\xi_c(t_{k+1})&=A_c\xi_c(t_k)+B_c\hat{y}(t_k)\\
v(t_k)&=C_c\xi_c(t_k)+D_c\hat{y}(t_{k}),
\end{aligned}
\right.
\end{equation}
where $\xi_c(t_k)\in\mathbb{R}^{n_c}$, $v(t_k)\in\mathbb{R}^{n_v}$, and $\hat{y}(t_k)\in\mathbb{R}^{n_y}$ denote the state vector, output vector of the controller, and input applied to the controller respectively. Define $h>0$ the sampling interval. A periodic sampling sequence is given by:
\begin{equation*}
\mathcal{T}:=\{t_k|t_k:=kh,k\in\mathbb{N}\}.
\end{equation*}
Define $\tau(t)$ be the elapsed time since the last sampling time, i.e. $\tau(t):=t-t_k,\,t\in[t_k,t_{k+1}[$. Define two vectors for the implementation input and output $u(t):=[y^{\mathrm{T}}(t)\, v^{\mathrm{T}}(t) ]^{\mathrm{T}}\in\mathbb{R}^{n_u}$, $\hat{u}(t_k):=[\hat{y}^{\mathrm{T}}(t_k)\, \hat{v}^{\mathrm{T}}(t_k)]^{\mathrm{T}}\in\mathbb{R}^{n_u}$, with $n_u:=n_y+n_v$. $u^i(t_k)$ $\hat{u}^i(t_k)$ are the $i$-th elements of the vector $u(t_k)$, $\hat{u}(t_k)$ respectively. At each sampling time $t_k\in\mathcal{T}$, the input applied to the implementation $\hat{u}(t_k)$ is determined by:
\begin{equation}\label{eq:sampleandupdateu}
\hat{u}^i(t_k):=\left\{\begin{aligned}
&\tilde{q}(u^i(t_k)),\,\text{if a local event triggered}\\
&\hat{u}^i(t_{k-1}),\,\text{otherwise},
\end{aligned}\right.
\end{equation}
where $\tilde{q}(s)$ denotes the quantized signal of $s$. Therefore, at each sampling time, only those inputs that triggered events are required to transmit measurements or actuation signals through the network. Between samplings, a zero-order hold mechanism is applied. We also introduce a performance variable $z\in\mathbb{R}^{n_z}$ given by:
\begin{equation}\label{eq:z}
z(t)=g(\xi(t),w(t)),
\end{equation}
where $\xi(t):=[\xi_p^{\mathrm{T}}(t)\,\xi_c^{\mathrm{T}}(t)\,
\hat{y}^{\mathrm{T}}(t) \,\hat{v}^{\mathrm{T}}(t)]^{\mathrm{T}}\in\mathbb{R}^{n_{\xi}}$, $n_{\xi}:=n_p+n_c+n_y+n_v$, and $g(s)$ is a design function. In this implementation, the controller, sensors, and actuators are assumed to be physically distributed, and none of the nodes are co-located. We employ the definition of uniform global pre-asymptotic stable (UGpAS), Lyapunov function candidate, and sufficient Lyapunov conditions for UGpAS from \cite{goebel2009hybrid}.

\begin{defn}{($\mathcal{L}_2$-gain)\cite{heemels2013periodic}}\label{defi:l2gain}
The system (\ref{eq:system}), (\ref{eq:controllaw}), (\ref{eq:z}) is said to have an $\mathcal{L}_2$-gain from $w$ to $z$ smaller than or equal to $\gamma$, if there is a $\mathcal{K}_{\infty}$ function $\delta:\mathbb{R}^{n_{\xi}}\rightarrow\mathbb{R}^+$ such that for any $w\in\mathcal{L}_2$, any initial state $\xi(0)=\xi_0\in\mathbb{R}^{n_{\xi}}$ and $\tau(0)\in[0,h]$, the corresponding solution to system (\ref{eq:system}), (\ref{eq:controllaw}), (\ref{eq:z}) satisfies $\|z\|_{\mathcal{L}_2}\leq\delta(\xi_0)+\gamma\|w\|_{\mathcal{L}_2}$.
\end{defn}

In the local event conditions in (\ref{eq:sampleandupdateu}), an event occurs when the following inequality holds:
\begin{equation}\label{eq:localevent}
\begin{aligned}
|\hat{u}^i(t_{k-1})-u^i(t_k)|\geq\sqrt{\eta_i(t_k)},\,i\in\{1,\cdots,n_{u}\},
\end{aligned}
\end{equation}
in which $\eta_i(t_k)$ is a local threshold, computed as:
\begin{equation}\label{eq:localthreshold}
\begin{aligned}
\eta_i(t):=\theta_i^2\eta^2(t),\\
\end{aligned}
\end{equation}
where $\theta_i$ is a designed distributed parameter satisfying $|\theta|=1$ and $\eta:\mathbb{R}^+_0\rightarrow\mathbb{R}^{+}$, determines the global threshold, which will be discussed in Section \ref{section:stabilityandperformance}. When an event takes place at a sampling time $t_k$, $\hat{u}(t_k)$ is updated by:
\begin{equation}\label{eq:update}
\begin{aligned}
&\hat{u}^i(t_k)=\tilde{q}(u^i(t_k))=q_{\eta}(u^i(t_k),\hat{u}^i(t_{k-1})):=\\
&\hat{u}^i(t_{k-1})-\text{sign}(\hat{u}^i(t_{k-1})-u^i(t_k))m^i(t_k)\sqrt{\eta_i(t_k)},\\
\end{aligned}
\end{equation}
where $m^i(t_k):=\left\lfloor\frac{|\hat{u}^i(t_{k-1})-u^i(t_k)|}{\sqrt{\eta_i(t_k)}}\right\rfloor$. The error after this update is:
\begin{equation}\label{eq:uerror}
\begin{aligned}
&e_u^i(t_k):=\hat{u}^i(t_k)-u^i(t_k)=-\text{sign}(\hat{u}^i(t_{k-1})-\\
&u^i(t_k))\left(m^i(t_k)-\frac{|\hat{u}^i(t_{k-1})- u^i(t_k)|}{\sqrt{\eta_i(t_k)}}\right)\sqrt{\eta_i(t_k)}.
\end{aligned}
\end{equation}
One can easily observe that, $|e_u^i(t_k)|<\sqrt{\eta_i(t_k)}$. That is, when there is an event locally, after the update by (\ref{eq:update}), (\ref{eq:localevent}) does not hold anymore. Later we show that, $\forall i\in\{1,\cdots,n_u\},\,k\in\mathbb{N}, \,m^i(t_k)\leq\bar{m}_{x}<\infty$. Thus, in practice one only needs to send $\text{sign}(\hat{u}^i(t_{k-1})-u^i(t_k))$ and $m^i(t_k)$ for each input update. Therefore, only $\log_2 (m^i(t_k))+1$ bits are required for each transmission from a single sensor or to a single actuator. Define $\Gamma_{\mathcal{J}}:=\text{diag}(\Gamma^y_{\mathcal{J}},\Gamma^v_{\mathcal{J}})=\text{diag}(\gamma^1_{\mathcal{J}} \cdots,\gamma_{\mathcal{J}}^{n_u})$, where $\mathcal{J}$ is an index set: $\mathcal{J}\subseteq\bar{\mathcal{J}}=\{1,\cdots,n_u\}$ for $u(t)$, indicating the occurrence of events. Define $\mathcal{J}_c:=\bar{\mathcal{J}}\setminus\mathcal{J}$.  For $l\in\{1,\cdots,n_u\}$, if $l\in\mathcal{J}$, $\gamma_{\mathcal{J}}^l=1$; if $l\in\mathcal{J}_c$, $\gamma_{\mathcal{J}}^l=0$. Furthermore, we use the notation $\Gamma_j=\Gamma_{\{j\}}$. Define $C:=\begin{bmatrix}
                                                  C_p & 0 \\
                                                  0 & C_c \\
                                                \end{bmatrix}$ and $D:=\begin{bmatrix}
                                                  0 & 0 \\
                                                  D_c & 0 \\
                                                \end{bmatrix}$.
The local event-triggered condition (\ref{eq:localevent}) can now be reformulated as a set membership:
\begin{equation}\label{eq:eventcondition}
\begin{aligned}
&i\in\mathcal{J}\text{ iff }\xi^{\mathrm{T}}(t_k)Q_i\xi(t_k)\geq\eta_i(t_k),
\end{aligned}
\end{equation}
where $Q_i=\begin{bmatrix}
             C^{\mathrm{T}}\Gamma_iC & C^{\mathrm{T}}\Gamma_iD-C^{\mathrm{T}}\Gamma_i \\
             D^{\mathrm{T}}\Gamma_iC-\Gamma_iC & (D-I)^{\mathrm{T}}\Gamma_i(D-I) \\
\end{bmatrix}$. The ADPETC implementation determined by (\ref{eq:system}), (\ref{eq:controllaw}), (\ref{eq:sampleandupdateu}), (\ref{eq:z}), and (\ref{eq:eventcondition}) can be re-written as an impulsive system model:
\begin{equation}\label{eq:impulsivesystem}
\begin{aligned}
\begin{bmatrix}
  \dot{\xi}(t) \\
  \dot\tau(t) \\
\end{bmatrix}&=\begin{bmatrix}
                 \bar{A}\xi(t)+\bar{B}w(t) \\
                 1 \\
               \end{bmatrix},\,\text{when }\tau(t)\in[0,h[,\\
\begin{bmatrix}
  \xi(t_k^+) \\
  \tau(t_k^+) \\
\end{bmatrix}&=\begin{bmatrix}
                 J_{\mathcal{J}}\xi(t_k)+\Delta_{\mathcal{J}}(t_k)\eta(t_k) \\
                 0 \\
               \end{bmatrix},\,\text{when }\tau(t)=h,\\
z(t)&=g(\xi(t),w(t)),
\end{aligned}
\end{equation}
where $\bar{B}=\begin{bmatrix}
          E^{\mathrm{T}} & 0 & 0 & 0 \\
        \end{bmatrix}^{\mathrm{T}}$ and
\begin{equation*}
\begin{aligned}
\bar{A}&=\begin{bmatrix}
          A_p & 0 & 0 & B_p \\
          0 & 0 & 0 & 0 \\
          0 & 0 & 0 & 0 \\
          0 & 0 & 0 & 0 \\
        \end{bmatrix},\,\Delta_{\mathcal{J}}(t_k)=\begin{bmatrix}
          0 \\
          B_c\Gamma^y_{\mathcal{J}}\epsilon_y(t_k)\Theta_y \\
          \Gamma^y_{\mathcal{J}}\epsilon_y(t_k)\Theta_y \\
          \Gamma^v_{\mathcal{J}}\epsilon_v(t_k)\Theta_v \\
        \end{bmatrix},\\
J_{\mathcal{J}}&=\begin{bmatrix}
                   I & 0 & 0 & 0 \\
                   B_c\Gamma^y_{\mathcal{J}}C_p & A_c & B_c(I-\Gamma^y_{\mathcal{J}}) & 0 \\
                   \Gamma^y_{\mathcal{J}}C_p & 0 & (I-\Gamma^y_{\mathcal{J}}) & 0 \\
                   0 & \Gamma^v_{\mathcal{J}}C_c & \Gamma_{\mathcal{J}}^vD_c & (I-\Gamma^v_{\mathcal{J}}) \\
                 \end{bmatrix},
\end{aligned}
\end{equation*}
with $I$ an identity matrix of corresponding dimension,
\begin{equation*}
\begin{aligned}
&\epsilon_y(t_k):=\text{diag}\left(\frac{e^1_u(t_k)}{\sqrt{\eta_1(t_k)}},\cdots, \frac{e^{n_y}_u(t_k)}{\sqrt{\eta_{n_y}(t_k)}}\right),\\
&\epsilon_v(t_k):=\text{diag}\left(\frac{e^{n_y+1}_u(t_k)}{\sqrt{\eta_{n_y+1}(t_k)}},\cdots, \frac{e^{n_y+n_v}_u(t_k)}{\sqrt{\eta_{n_y+n_v}(t_k)}}\right),\\
&\Theta_y:=\begin{bmatrix}
\theta_1 & \cdots & \theta_{n_y} \\
\end{bmatrix}^{\mathrm{T}},\,
\Theta_v:=\begin{bmatrix}
\theta_{n_y+1} & \cdots & \theta_{n_y+n_v} \\
\end{bmatrix}^{\mathrm{T}}.
\end{aligned}
\end{equation*}
The term $\Delta_{\mathcal{J}}(t_k)\eta(t_k)$ represents the quantization error after input updates and $\frac{e^i_u(t_k)}{\sqrt{\eta_{i}(t_k)}}\in]-1,1[$ due to (\ref{eq:update}), (\ref{eq:uerror}). Lemma 9 in \cite{MazoJr2014} indicates that, for a system applying the ADETC mechanism to be uniformly globally asymptotically stable (UGAS, see \cite{MazoJr2014}
) when $w=0$, $\eta(t)$ should be a monotonically decreasing function with $\lim_{t\rightarrow\infty}\eta(t)=0$. However, this mechanism does not consider systems with disturbances. According to \cite{liberzon2007input}, when $w\neq 0$, if $\eta(t)$ is arbitrarily small, the mechanism is not robust against disturbances. Meanwhile, in \cite{MazoJr2014}, the $\eta(t)$ update is determined by an upper bound estimate of the current state of the plant. This estimate is not always obtainable in an output-feedback system, making it unapplicable in such systems. We overcome the first problem by imposing a lower bound on $\eta(t_k)$, defined as $\eta_{\min}>0$, i.e. $\eta(t_k)\geq\eta_{\min},\forall t_k\in\mathcal{T}$. For the second problem, we instead use $\xi_c(t_k)$, $\hat{y}(t_k)$, and $\hat{v}(t_k)$ to determine the current threshold instead of $\xi_p(t_k)$, since this information is available to the controller.

\begin{rem}
By imposing a lower bound $\eta_{\min}$ on $\eta$, the $\lim_{t\rightarrow\infty}\eta(t)\neq0$, and thus $\xi(t)$ can only converge to a set even when $w=0$. Therefore, no $\mathcal{L}_{2}$-gain can be obtained for a linear performance function, proportional to the state of the system as in \cite{heemels2013periodic}, since in that case $\xi\notin\mathcal{L}_2$ implies $z\notin\mathcal{L}_2$. We circumvent this problem picking a performance function that is zero on a compact set around the origin.
\end{rem}

Denote the solution set $\mathcal{X}$ as $(x,r)\in\mathcal{X}\subseteq\mathbb{R}^{n_{\xi}}\times[0,h]$, such that $x=\xi(t)$, $r=\tau(t)$ for some $t\in\mathbb{R}^+_0$, where $\xi$ is a solution to system (\ref{eq:impulsivesystem}). $\mathcal{A}\subseteq\mathcal{X}$ is a compact set around the origin. Re-define the variable $z(t)$ in (\ref{eq:impulsivesystem}) by:
\begin{equation}\label{eq:newz}
z_{\mathcal{A}}(t):=\left\{
\begin{aligned}
&\bar{C}\xi(t)+\bar{D}w(t),\,\forall (\xi(t),\tau(t))\in\mathcal{X}\setminus\mathcal{A}\\
&0,\,\forall (\xi(t),\tau(t))\in\mathcal{A},
\end{aligned}\right.
\end{equation}
in which, $\bar{C}$ and $\bar{D}$ are some matrices of appropriate dimensions. Now we present the main problem we solve in this paper.

\begin{prob}
Design an update mechanism for $\eta$ and an $\eta_{\min}$ such that $\mathcal{A}$ is UGpAS for (\ref{eq:impulsivesystem}), (\ref{eq:newz}) when $w=0$, and the $\mathcal{L}_2$-gain from $w$ to $z_{\mathcal{A}}$ is smaller than or equal to $\gamma$.
\end{prob}

\section{Stability and $\mathcal{L}_2$-gain analysis}
\label{section:stabilityandperformance}

Denote $\tilde{z}(t)$ a reference function of $z_{\mathcal{A}}(t)$, given by:
\begin{equation}\label{eq:tildez}
\tilde{z}(t):=\bar{C}\xi(t)+\bar{D}w(t),\,\forall (\xi(t),\tau(t))\in\mathcal{X}.
\end{equation}
Now let us consider a Lyapunov function candidate for the impulsive system (\ref{eq:impulsivesystem}), (\ref{eq:tildez}) of the form:
\begin{equation}\label{eq:lyapunov}
V(x,r)=x^{\mathrm{T}}P(r)x,
\end{equation}
where $x\in\mathbb{R}^{n_\xi}$, $r\in[0,h]$, with $P:[0,h]\rightarrow\mathbb{R}^{n_\xi\times n_{\xi}}$ satisfying the Riccati differential equation:
\begin{equation}\label{eq:P}
\frac{d}{d r}P=-\bar{A}^{\mathrm{T}}P-P\bar{A}-2\rho P-\gamma^{-2}\bar{C}^{\mathrm{T}}\bar{C}-G^{\mathrm{T}}MG,
\end{equation}
in which $M:=(I-\gamma^{-2}\bar{D}^{\mathrm{T}}\bar{D})^{-1}$; $G:=\bar{B}^{\mathrm{T}}P+\gamma^{-2}\bar{D}^{\mathrm{T}}\bar{C}$, with $\bar{A}$, $\bar{B}$, $\bar{C}$, and $\bar{D}$ defined in (\ref{eq:impulsivesystem}) and (\ref{eq:tildez}), and $\rho$ and $\gamma$ are pre-design parameters. We often use the shorthand notation $V(t)$ to denote $V(\xi(t),\tau(t))$. Construct the Hamiltonian matrix:
\begin{equation*}\label{eq:H}
\begin{aligned}
H:=\begin{bmatrix}
     H_{11} & H_{12} \\
     H_{21} & H_{22} \\
   \end{bmatrix},\,F(r):=e^{-Hr}=\begin{bmatrix}
                      F_{11}(r) & F_{12}(r) \\
                      F_{21}(r) & F_{22}(r) \\
                    \end{bmatrix},
\end{aligned}
\end{equation*}
where $
H_{11}:=\bar{A}+\rho I+\gamma^{-2}\bar{B}M\bar{D}^{\mathrm{T}}\bar{C},\, H_{12}:=\bar{B}M\bar{B}^{\mathrm{T}},\, H_{21}:=-\bar{C}^{\mathrm{T}}(\gamma^{2}I-\bar{D}\bar{D}^{\mathrm{T}})^{-1}\bar{C},\, H_{22}:=-(\bar{A}+\rho I+\gamma^{-2}\bar{B}M\bar{D}^{\mathrm{T}}\bar{C})^{\mathrm{T}}$.
\begin{assum}\label{assumption}
$F_{11}(r)$ is invertible $\forall r\in[0,h]$.
\end{assum}
Since $F_{11}(0)=I$ and $F_{11}(r)$ is continuous, Assumption \ref{assumption} can always be satisfied for sufficiently small $h$. According to Lemma A.1 in \cite{heemels2013periodic}, if Assumption \ref{assumption} holds, then $-F_{11}^{-1}(h) F_{12}(h)$ is positive semi-definite. Define the matrix $\bar{S}$ satisfying $\bar{S}\bar{S}^{\mathrm{T}}:=-F_{11}^{-1}(h)F_{12}(h)$.

We present next the designed threshold update mechanism. At each sampling time $t_k^+$, right after a jump of system (\ref{eq:impulsivesystem}), the controller executes the threshold update mechanism:
\begin{equation}\label{eq:thresholdupdate}
\begin{aligned}
\eta(t_k^+)&=\mu^{-n_{\mu}(t_k^+)}\eta_{\min},
\end{aligned}
\end{equation}
in which $n_{\mu}(t_k^+):=\max\left\{0,\left\lceil-\log_{\mu}\left(\frac{|\xi'(t_k^+)|} {\varrho\eta_{\min}}\right)-1\right\rceil\right\}$, $\eta_{\min}$ is a pre-designed minimum threshold; finite $\varrho>0$ is a design parameter; and the scalar $\mu:\in]0,1[$ is also a pre-designed parameter. The vector of variables available at the controller at sampling time $t_k^+$ is denoted by $\xi'(t_k^+):=[\xi_c^{\mathrm{T}}(t_k^+)\, \hat{y}^{\mathrm{T}}(t_k^+)\, \hat{v}^{\mathrm{T}}(t_k^+)]^{\mathrm{T}}$.
\begin{lem}\label{lemma:threshold}
Consider the system (\ref{eq:impulsivesystem}), (\ref{eq:tildez}), after the execution of the threshold update mechanism (\ref{eq:thresholdupdate}), if $\eta(t_k^+)\neq\eta_{\min}$, then: $\varrho\eta(t_k^+)<|\xi'(t_k^+)|\leq\mu^{-1}\varrho\eta(t_k^+)$.
\end{lem}
Now we analyze the jump part of the impulsive system.

\begin{lem}\label{lemma:jump2}
Consider the system (\ref{eq:impulsivesystem}), (\ref{eq:tildez}), (\ref{eq:lyapunov}), (\ref{eq:P}), and (\ref{eq:thresholdupdate}), and that Assumption \ref{assumption} holds. If $\gamma^2>\lambda_{\max}(\bar{D}^{\mathrm{T}}\bar{D})$, $\exists P(h)\succ0$ satisfying $I-\bar{S}^{\mathrm{T}}P(h)\bar{S}\succ 0$, and scalars $\varrho>0$, $\epsilon>0$ such that the LMI:
\begin{equation}\label{eq:jump2}
\begin{aligned}
\begin{bmatrix}
  \epsilon I & \tilde{F}_1 & \tilde{F}_2 & -\epsilon J_{\mathcal{\bar{J}}} \\
  \tilde{F}_1^{\mathrm{T}} & \tilde{F}_3 & 0 & 0 \\
  \tilde{F}_2^{\mathrm{T}} & 0 & \tilde{F}_2 & 0 \\
  -\epsilon J_{\mathcal{\bar{J}}}^{\mathrm{T}} & 0 & 0 & P(h)+\epsilon J_{\mathcal{\bar{J}}}^{\mathrm{T}}J_{\mathcal{\bar{J}}}-\epsilon\frac{|\bar\Delta_{\mathcal{\bar{J}}}|^2} {\varrho^2}I \\
\end{bmatrix}\succeq 0
\end{aligned}
\end{equation}
holds, where $
\tilde{F}_1:=F_{11}^{-\mathrm{T}}(h)P(h)\bar{S}$, $\tilde{F}_3:=I-\bar{S}^{\mathrm{T}}P(h)\bar{S}$, $
\tilde{F}_2:=F_{11}^{-\mathrm{T}}(h)P(h)F_{11}^{-1}(h)+F_{21}(h)F_{11}^{-1}(h)$, $
\bar\Delta_{\mathcal{J}}:=\Delta_{\mathcal{J}}(t_k)_{|\epsilon_y(t_k)=I,\epsilon_v(t_k)=I}$,
then $\forall t_k\in\mathcal{T}$ such that $|\xi(t_k)|>\varrho\eta(t_k)$, the following also holds: $V(\xi(t_k^+),0)\leq V(\xi(t_k),h)$.
\end{lem}

Note that $\varrho$ enters the LMI in a nonlinear fashion, therefore we cannot compute $\varrho$ directly. Instead, we apply a line search algorithm to find feasible parameters $h$ and $\varrho$. Define $C_H=\{(x,r)|(x,r)\in\mathcal{X},r\in[0,h[\}$, $D_H=\{(x,r)|(x,r)\in\mathcal{X},r=h\}$, and the set $\underline{\mathcal{A}}$ as:
\begin{equation}\label{eq:a}
\begin{aligned}
&\underline{\mathcal{A}}:=\left\{(x,r)|(x,r)\in\mathcal{X},V(x,r)\leq\bar\lambda\bar\varrho^2\eta_{\min}^2\right\},
\end{aligned}
\end{equation}
where $\bar\lambda:=\max\{\lambda_{\max}(P(r)),\allowbreak\forall r\in[0,h]\}$, $\bar\varrho:=\max\{|J_{\mathcal{J}}|\varrho+|\bar\Delta_{\mathcal{J}}|,\forall \mathcal{J}\subseteq\bar{\mathcal{J}}\}$. Selecting $\eta_{\min}$ sufficiently small, one can make sure that $\underline{\mathcal{A}}\subseteq\mathcal{A}$. Define now a new Lyapunov function candidate for system (\ref{eq:impulsivesystem}), (\ref{eq:tildez}), and (\ref{eq:thresholdupdate}), as:
\begin{equation}\label{eq:newlpf}
\begin{aligned}
W(x,r):=\max\{V(x,r)-\bar\lambda\bar\varrho^2\eta_{\min}^2,0\}.\\
\end{aligned}
\end{equation}
Note that (\ref{eq:newlpf}) defines a proper Lyapunov function candidate. We also use the shorthand notation $W(t)$ to denote $W(\xi(t),\tau(t))$. Finally, let:
\begin{equation}\label{eq:nnewz}
z_{\underline{\mathcal{A}}}(t):=\left\{
\begin{aligned}
&\bar{C}\xi(t)+\bar{D}w(t),\,\forall (\xi(t),\tau(t))\in\mathcal{X}\setminus\underline{\mathcal{A}}\\
&0,\,\forall (\xi(t),\tau(t))\in\underline{\mathcal{A}}.
\end{aligned}\right.
\end{equation}
It is obvious that if $\underline{\mathcal{A}}\subseteq\mathcal{A}$, $|z_{\underline{\mathcal{A}}}(t)|\geq|z_{\mathcal{A}}(t)|\geq 0$.

\begin{thm}\label{theorem:stabilityandperformance}
Consider the system (\ref{eq:impulsivesystem}), (\ref{eq:newz}), (\ref{eq:lyapunov}), (\ref{eq:P}), (\ref{eq:thresholdupdate}), (\ref{eq:a}), and (\ref{eq:newlpf}). If $\rho>0$, $\gamma^2>\lambda_{\max}(\bar{D}^{\mathrm{T}}\bar{D})$, the hypotheses of Lemma \ref{lemma:jump2} hold, and $\eta_{\min}$ is selected s.t. $\underline{\mathcal{A}}\subseteq\mathcal{A}$, then $\mathcal{A}$ is UGpAS for the impulsive system (\ref{eq:impulsivesystem}) when $w=0$; and the $\mathcal{L}_2$-gain from $w$ to $z_{\mathcal{A}}$ is smaller than or equal to $\gamma$.
\end{thm}

\section{Practical considerations}

In our proposed implementation, the data a sensor sends is actually $m^i(t_k)$ and the sign of the error, see (\ref{eq:update}). Therefore, computing an upper bound $\bar{m}_x\geq m^i(t_k)$, $\forall t_k\in\mathcal{T}$ is desirable to properly design the supporting communication protocol.

\begin{prop}\label{lemma:mxupperbound}
Consider the system (\ref{eq:impulsivesystem}), (\ref{eq:newz}), (\ref{eq:lyapunov}), (\ref{eq:P}), (\ref{eq:thresholdupdate}), and (\ref{eq:newlpf}). If $w$ is bounded (i.e. $w\in\mathcal{L}_2\cap\mathcal{L}_{\infty}$), and the hypotheses of Theorem \ref{theorem:stabilityandperformance} hold, then:
\begin{equation}\label{Eq:mxupperbound}
\begin{aligned}
\bar{m}_x&=\max\{\bar{m}_x^i|i\in\{1,\cdots,n_u\}\}\\
\end{aligned}
\end{equation}
where $\bar{m}_x^i=\frac{(1+|[C\,D]|)}{\theta_i}\sqrt{\frac{W(0)}{\eta_{\min}^2\underline{\lambda}}+ \frac{\|w\|_{\mathcal{L}_{\infty}}^2}{2\rho\eta_{\min}^2\underline{\lambda}}+ \frac{\bar\lambda\bar\varrho^2}{\underline{\lambda}}}\geq m^i(t_k),\,\forall t_k\in\mathcal{T}$; $\underline\lambda=\min\{\lambda_{\min}(P(r)),\forall r\in[0,h]\}$.
\end{prop}

Similarly, an upper bound of $n_{\mu}(t)$, denoted by $\bar{m}_{\mu}$ can be obtained:
\begin{prop}\label{lemma:mmuupperbound}
Consider the system (\ref{eq:impulsivesystem}), (\ref{eq:newz}), (\ref{eq:lyapunov}), (\ref{eq:P}), (\ref{eq:thresholdupdate}), and (\ref{eq:newlpf}). If $w$ is bounded and the hypotheses of Theorem \ref{theorem:stabilityandperformance} hold, then $\bar{m}_{\mu}$ is given as $\bar{m}_{\mu} =\max\left\{0,-\log_{\mu}\left(\frac{(1+|[C\,D]|)}{\varrho}\sqrt{\frac{W(0)}{\eta_{\min}^2\underline{\lambda}}+ \frac{\|w\|_{\mathcal{L}_{\infty}}^2}{2\rho\eta_{\min}^2\underline{\lambda}}+ \frac{\bar\lambda\bar\varrho^2}{\underline{\lambda}}}\right)\right\}$.
\end{prop}




\section{Numerical example}

\begin {figure}[!t]
\centering
\includegraphics[width=\linewidth]{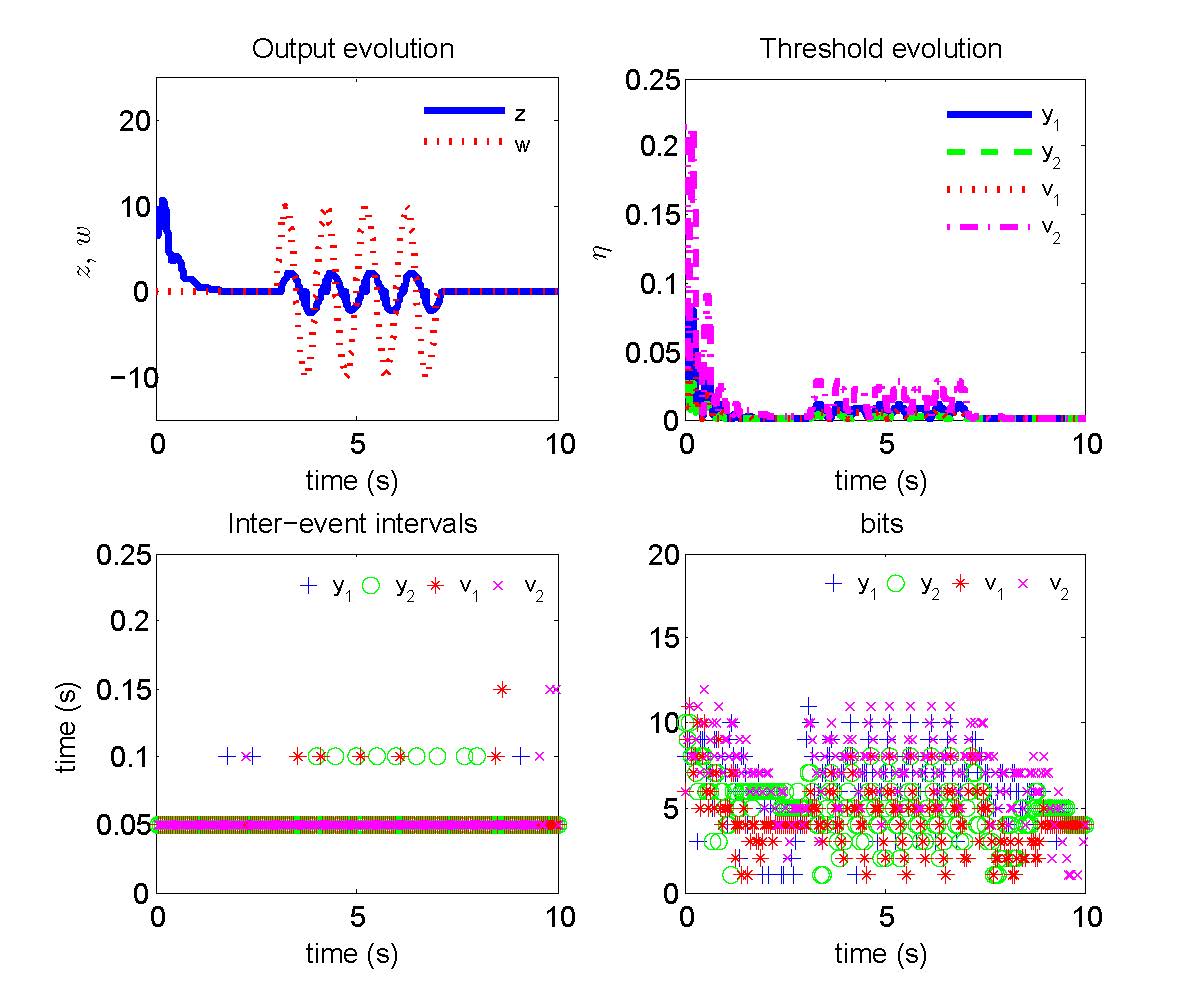}
\caption{Simulation result when $w(t)=10\sin(2\pi t)$, $t=[3,7]$: evolution of $z$ and $w$, threshold, inter-event intervals, and bits of each event.}
\label{fig:disturbance}
\end {figure}

In this section, we consider the batch reactor system from \cite{walsh2001scheduling}. Given $h=0.05s$,
with $\rho=0.01$, $\gamma=0.9$, $z=[1\,0\,0\,0\,0\,0\,0\,0\,0\,0]\xi$, $\mathcal{A}=\{(x,r)|(x,r)\in\mathcal{X},|x^{\mathrm{T}}P(r)x|\leq 3.11\}$. Assumption \ref{assumption} is satisfied. Solving (\ref{eq:jump2}), one can obtain a $\varrho=200.2$. Other parameters are given by $\mu=0.75$, $\theta_1= 0.34$, $\theta_2=0.11$, $\theta_3=0.23$, and $\theta_4=0.91$. $\xi_p(0)=[10\,-10\,-10\,10]^{\mathrm{T}}$, $\xi_c(0)=\textbf{0}$, $\hat{y}(0)=C_p\xi_p(0)$, and $\hat{v}(0)=D_cC_p\xi_p(0)$. Let $\eta_{\min}=0.0001$,
resulting in the set $\underline{\mathcal{A}}=\mathcal{A}$. Fig \ref{fig:disturbance} shows the simulation results in the presence of a finite sine wave disturbance. It can be seen that the performance variable $z$ follows $w$ with a bounded norm ratio. The sensor transmissions are reduced by $3.61\%$ compared to a time-triggered mechanism with the same sampling interval $h$. The maximum inter-event interval is 0.15 seconds. The following bounds are obtained from our analysis: $\bar{m}_x=2.40\times 10^8$ (29 bits), and $\bar{m}_{\mu}=42$. $89.81\%$ of $m^i(t_k)$ are smaller than or equal to 128 (8 bits); $31.23\%$ of $m^i(t_k)$ can be transmitted with 4 bits; and the maximum $m^i(t_k)$ is 1303 (12 bits). Note that the saving of transmission increases as the time without disturbances increases. Further simulation results show that, the sensor transmissions are reduced by $63.81\%$ after running for $50s$ without additional disturbances. Further simulation also shows that, as the initial state is closer to the original point, the reduction within $10$ seconds increases when there is no disturbance. When there are disturbances, the reduction does not change much.


\section{Conclusion and future work}

We propose ADPETC implementations as an extension to the work of \cite{heemels2013periodic} and \cite{MazoJr2014}. This triggering strategy combines decentralized event generation, asynchronous sampling update, and zoom in/out quantization. This approach lets the implementation exchange very few bits every time that an event triggers a transmission, reduces the required amount of transmission compared to time-triggered mechanisms, and reduces the necessary sensing compared to continuously monitored event-triggered mechanisms. The maximum amounts of bits that may be needed to update samplings and thresholds after an event is triggered are provided.
Such a bound enables the design of actual implementations for wireless systems, whose demonstration on physical experiments is part of our future work.
How to optimize $\mu$ and how to compensate transmission delays are additional goals for future work.


\section*{Appendix. Proofs}

The following two lemmas are intermediate results from the proof of Theorem III.2 in \cite{heemels2013periodic}, which will be used in the proofs of Lemma \ref{lemma:jump2} and Theorem \ref{theorem:stabilityandperformance}.\\
\textbf{Lemma 10} Consider the system (\ref{eq:impulsivesystem}), (\ref{eq:tildez}), (\ref{eq:lyapunov}), (\ref{eq:P}), and that Assumption \ref{assumption} holds. If $\gamma^2>\lambda_{\max}(\bar{D}^{\mathrm{T}}\bar{D})$ and $\exists P(h)>0$ satisfying $I-\bar{S}^{\mathrm{T}}P(h)\bar{S}\succ 0$, then for $\tau(t)\in[0,h]$, $P(\tau(t))\succ 0$; and $P(0)$ can be expressed as $P(0)=F_{21}(h)F_{11}^{-1}(h)+F_{11}^{-\mathrm{T}}(h)(P(h)+ P(h)\bar{S}(I-\bar{S}^{\mathrm{T}}P(h)\bar{S})^{-1}\bar{S}^{\mathrm{T}}P(h))F_{11}^{-1}(h)$.\\
\textbf{Lemma 11} Consider the system (\ref{eq:impulsivesystem}), (\ref{eq:tildez}), (\ref{eq:lyapunov}), and (\ref{eq:P}). If $\rho>0$, $\gamma^2>\lambda_{\max}(\bar{D}^{\mathrm{T}}\bar{D})$, then for all $x\in\mathbb{R}^{n_{\xi}}$ and $\tau(t)\in[0,h]$, the following inequation holds: $\frac{d}{dt}V(t)\leq-2\rho V(t)-\gamma^{-2}\tilde{z}^{\mathrm{T}}(t)\tilde{z}(t)+w^{\mathrm{T}}(t)w(t)$.\\
\textbf{Proof of Lemma \ref{lemma:threshold}}
For any $s=\left\lceil-\log_{\mu}(\frac{|\xi'(t_k^+)|}{\varrho\eta_{\min}})-1\right\rceil$, $s$ satisfies $-\log_{\mu}\left(\frac{|\xi'(t_k^+)|}{\varrho\eta_{\min}}\right)-1\leq s<-\log_{\mu}\left(\frac{|\xi'(t_k^+)|} {\varrho\eta_{\min}}\right)$. Noting that $\mu\in]0,1[$, therefore it is easy to obtain that $\mu^{\log_{\mu}\left(\frac{|\xi'(t_k^+)|}{\varrho\eta_{\min}}\right)+1}\leq \mu^{-s}<\mu^{\log_{\mu}\left(\frac{|\xi'(t_k^+)|} {\varrho\eta_{\min}}\right)}$, which, as $\varrho\eta_{\min}>0$, can be finally simplified as $\mu|\xi'(t_k^+)|\leq \varrho\mu^{-s}\eta_{\min}<|\xi'(t_k^+)|$. From (\ref{eq:thresholdupdate}), after the execution of the threshold update mechanism, $\eta(t_k^+)$ can be computed as $\eta(t_k^+)=\max\{\eta_{\min},\mu^{-s}\eta_{\min}\}$. If $\eta(t_k^+)\neq\eta_{\min}$, then $\eta(t_k^+)=\mu^{-s}\eta_{\min}$, and thus we have that $\mu|\xi'(t_k^+)|\leq \varrho\eta(t_k^+)<|\xi'(t_k^+)|$.\qed \\
\textbf{Proof of Lemma \ref{lemma:jump2}}
For the jump part of the impulsive system (\ref{eq:impulsivesystem}), we have that the relation between the states before and after each jump is given by $|\xi(t_k^+)-J_{\mathcal{\bar{J}}}\xi(t_k)|=|J_{\mathcal{J}}\xi(t_k)+ \Delta_{\mathcal{J}}(t_k)\eta(t_k)-J_{\mathcal{\bar{J}}}\xi(t_k)|
=|\tilde{H}_1\xi(t_k)+\Delta_{\mathcal{J}}(t_k)\eta(t_k)|$, where $\tilde{H}_1:=\begin{bmatrix}
0 & 0 & 0 & 0 \\
-B_c\Gamma^y_{\mathcal{J}_c}C_p & 0 & B_c\Gamma^y_{\mathcal{J}_c} & 0 \\
-\Gamma^y_{\mathcal{J}_c}C_p & 0 & \Gamma^y_{\mathcal{J}_c} & 0 \\
0 & -\Gamma^v_{\mathcal{J}_c}C_c & -\Gamma_{\mathcal{J}_c}^vD_c & \Gamma^v_{\mathcal{J}_c} \\
\end{bmatrix}$, since $\Gamma_{\mathcal{J}_c}^y+\Gamma_{\mathcal{J}}^y=I=\Gamma_{\mathcal{\bar{J}}}^y$ and $\Gamma_{\mathcal{J}_c}^v+\Gamma_{\mathcal{J}}^v=I=\Gamma_{\mathcal{\bar{J}}}^v$. By the definition of error (\ref{eq:uerror}) and  the event-triggered mechanism (\ref{eq:eventcondition}), one has $\Gamma_{\mathcal{J}_c}^y\hat{y}(t_k)-\Gamma_{\mathcal{J}_c}^yy(t_k)= \Gamma_{\mathcal{J}_c}^y\epsilon_y(t_k)\Theta_y\eta(t_k)$ and $\Gamma_{\mathcal{J}_c}^v\hat{v}(t_k)-\Gamma_{\mathcal{J}_c}^vv(t_k)= \Gamma_{\mathcal{J}_c}^v\epsilon_v(t_k)\Theta_v\eta(t_k)$, therefore, it holds that $\tilde{H}_1\xi(t_k)+\Delta_{\mathcal{J}}(t_k)\eta(t_k)= \Delta_{\mathcal{J}_c}(t_k)\eta(t_k)+\Delta_{\mathcal{J}}(t_k)\eta(t_k) =\Delta_{\mathcal{\bar{J}}}(t_k)\eta(t_k)$, and thus $|\xi(t_k^+)-J_{\mathcal{\bar{J}}}\xi(t_k)|= |\Delta_{\mathcal{\bar{J}}}(t_k)\eta(t_k)|\leq|\bar{\Delta}_{\mathcal{\bar{J}}}|\eta(t_k)$. Together with the hypothesis that $|\xi(t_k)|>\varrho\eta(t_k)$, one has $|(\xi(t_k^+)-J_{\mathcal{\bar{J}}}\xi(t_k))|^2< \frac{|\bar{\Delta}_{\mathcal{\bar{J}}}|^2}{\varrho^2}|\xi(t_k)|^2$. From the hypotheses, particularly (\ref{eq:jump2}) together with the result from Lemma 10, Schur complement, $\epsilon>0$, and applying the S-procedure, one can conclude that $V(\xi(t_k^+),0)\leq V(\xi(t_k),h)$.\qed \\
\textbf{Proof of Theorem \ref{theorem:stabilityandperformance}}
We first show that $\mathcal{A}$ is UGpAS for the impulsive system (\ref{eq:impulsivesystem}) when $w=0$. A new Lyapunov function candidate $W$, given by (\ref{eq:newlpf}), is introduced. Define $\mathcal{B}:=\{(x,r)|(x,r)\in\mathcal{X},|x|\leq\varrho\eta_{\min}\}$. If $\eta(t_k)=\eta_{\min}$, $|\xi(t_k)|>\varrho\eta_{\min}$ implies $|\xi(t_k)|>\varrho\eta(t_k)$; if $\eta(t_k)>\eta_{\min}$, according to Lemma \ref{lemma:threshold}, $\varrho\eta(t_k)<|\xi'(t_k)|\leq|\xi(t_k)|$. Therefore, $\forall (\xi(t_k),\tau(t_k))\in D_H\setminus\mathcal{B}$, $|\xi(t_k)|>\varrho\eta(t_k)$, and thus from Lemma \ref{lemma:jump2}, $\forall (\xi(t_k),\tau(t_k))\in D_H\setminus\mathcal{B}$, it holds that $V(\xi(t_k^{+}),0)\leq V(\xi(t_k),h)$. According to Lemma \ref{lemma:threshold}, if $|\xi'(t_k)|\leq\varrho\eta(t_k)$ then $\eta(t_k)=\eta_{\min}$, i.e. $\forall (\xi(t_k),\tau(t_k))\in D_H\cap\mathcal{B}$, $\eta(t_k)=\eta_{\min}$. Furthermore, $(\xi(t_k),\tau(t_k))\in D_H\cap\mathcal{B}$ implies $\xi(t_k^+)=J_{\mathcal{J}}\xi(t_k)+\Delta_{\mathcal{J}}\eta_{\min}$, and thus, $|\xi(t_k^+)|\leq|J_{\mathcal{J}}||\xi(t_k)|+|\Delta_{\mathcal{J}}|\eta_{\min}\leq(|J_{\mathcal{J}}| \varrho+|\bar\Delta_{\mathcal{J}}|)\eta_{\min}\leq\bar\varrho\eta_{\min}$. That is, $\forall (\xi(t_k),\tau(t_k))\in D_H\cap\mathcal{B}$, $(\xi(t^+_k),0)\in\underline{\mathcal{A}}$. Note that, since $|J_{\mathcal{J}}|>1$, $\forall (x,r)\in\mathcal{B}$, $x^{\mathrm{T}}P(r)x\leq\bar\lambda|x|^2\leq \bar\lambda\varrho^2\eta_{\min}^2<\bar\lambda\bar\varrho^2\eta_{\min}^2$, i.e. $\mathcal{B}\subset\underline{\mathcal{A}}$. Thus one can conclude that $\forall (\xi(t),\tau(t))\in \underline{\mathcal{A}}\cap D_H$, $(\xi(t^+_k),0)\in\underline{\mathcal{A}}$. If all the hypotheses in Lemma 11 hold, together with (\ref{eq:newlpf}), one has $\forall (\xi(t),\tau(t))\in C_H\setminus\underline{\mathcal{A}}$: $\frac{d}{dt}W(\xi(t),\tau(t))=\frac{d}{dt}V(\xi(t),\tau(t))\leq-2\rho V(\xi(t),\tau(t)) -\gamma^{-2}\tilde{z}^{\mathrm{T}}(t)\tilde{z}(t)+w^{\mathrm{T}}(t)w(t)<-2\rho W(\xi(t),\tau(t)) -\gamma^{-2}\allowbreak\tilde{z}^{\mathrm{T}}(t)\tilde{z}(t)+w^{\mathrm{T}}(t)w(t)$. By (\ref{eq:newlpf}) and $V(\xi(t_k^{+}),0)\leq V(\xi(t_k),h)$, one has $\forall(\xi(t_k),\tau(t_k))\in D_H\setminus\underline{\mathcal{A}}$: $W(\xi(t_k^{+}),0)=\max\{V(\xi(t_k^{+}),0)\allowbreak-\bar\lambda\bar\varrho^2\eta_{\min}^2,0\}\leq V(\xi(t_k),h)-\bar\lambda\bar\varrho^2\eta_{\min}^2=W(\xi(t_k),h)$. Combine all the above and $\underline{\mathcal{A}}\subseteq\mathcal{A}$ to see that $\mathcal{A}$ is UGpAS for the impulsive system (\ref{eq:impulsivesystem}).\\
Now we study the $\mathcal{L}_2$-gain. Define a set of times $\mathcal{T}_s=\{(t_i^s,j_i^s)|i\in\mathbb{N}\}$, where $(t_0^s,j_0^s)$ is the initial time, s.t. $\forall t\in[t_{2i+1}^s,t_{2i+2}^s]$, $i\in\mathbb{N}$, $(\xi(t),\tau(t))\in\underline{\mathcal{A}}$, and the rest of the time $(\xi(t),\tau(t))\in\mathcal{X}\setminus\underline{\mathcal{A}}$. If $|\mathcal{T}_s|$ is infinite, i.e. $(\xi(t),\tau(t))$ visits $\underline{\mathcal{A}}$ infinitely often, one has: $\int_{0}^{\infty}z_{\underline{\mathcal{A}}}^{\mathrm{T}}(t)z_{\underline{\mathcal{A}}}(t)dt= \sum_{i=0}^{\infty}\int_{t_i^s}^{t_{i+1}^s}z_{\underline{\mathcal{A}}}^{\mathrm{T}}(t)z_{\underline{\mathcal{A}}}(t)dt= \sum_{i=0}^{\infty}\int_{t_{2i}^s}^{t_{2i+1}^s}z_{\underline{\mathcal{A}}}^{\mathrm{T}}(t)z_{\underline{\mathcal{A}}}(t)dt+ \sum_{i=0}^{\infty}\int_{t_{2i+1}^s}^{t_{2i+2}^s}z_{\underline{\mathcal{A}}}^{\mathrm{T}}(t)z_{\underline{\mathcal{A}}}(t)dt$. $\forall (\xi(t),\tau(t))\in C_H\setminus\underline{\mathcal{A}}$, it holds that $\frac{d}{dt}W(\xi(t),\tau(t))<-\gamma^{-2}z_{\underline{\mathcal{A}}}^{\mathrm{T}}(t)z_{\underline{\mathcal{A}}}(t)+w^{\mathrm{T}}(t)w(t)$. One can replace the integration of $\frac{d}{dt}W(t)$, $z_{\underline{\mathcal{A}}}^{\mathrm{T}}(t)z_{\underline{\mathcal{A}}}(t)$, and $w^{\mathrm{T}}(t)w(t)$ on the open interval $]t_{2i}^s,t_{2i+1}^s[$ by the integration on the closure of that interval, see \cite{apostol1967calculus}. Applying the Comparison Lemma, one has $W(t_{2i+1}^s)-W(t_{2i}^s)=\int_{t_{2i}^s}^{t_{2i+1}^s}\frac{d}{dt}W(t)dt <\int_{t_{2i}^s}^{t_{2i+1}^s}\left(-\gamma^{-2}z_{\underline{\mathcal{A}}}^{\mathrm{T}}(t)z_{\underline{\mathcal{A}}}(t)+w^{\mathrm{T}}(t)w(t)\right)dt$. Since $\forall i\in\mathbb{N},i\neq0$, $W(t_i^s)=0$, therefore $\forall i\in\mathbb{N}$: $\sum_{i=0}^{\infty}\int_{t_{2i}^s}^{t_{2i+1}^s}\allowbreak z_{\underline{\mathcal{A}}}^{\mathrm{T}}(t)z_{\underline{\mathcal{A}}}(t)dt<  \gamma^2\sum_{i=0}^{\infty}\int_{t_{2i}^s}^{t_{2i+1}^s}w^{\mathrm{T}}(t)w(t)dt+ \gamma^2W(t_0^s)$. When $(\xi(t),\tau(t))\in\underline{\mathcal{A}}$, we have $z_{\underline{\mathcal{A}}}(t)=0$ from (\ref{eq:newz}), thus $
\sum_{i=0}^{\infty}\int_{t_{2i+1}^s}^{t_{2i+2}^s}z_{\underline{\mathcal{A}}}^{\mathrm{T}}(t)z_{\underline{\mathcal{A}}}(t)dt\leq\gamma^2 \sum_{i=0}^{\infty}\int_{t_{2i+1}^s}^{t_{2i+2}^s}w^{\mathrm{T}}(t)w(t)dt$. Combine all the above to obtain $\|z_{\mathcal{A}}\|_{\mathcal{L}_2}^2\leq\|z_{\underline{\mathcal{A}}}\|_{\mathcal{L}_2}^2<\gamma^2W(t_0^s)+\gamma^{2}\|w\|_{\mathcal{L}_2}^2 \leq\delta(\xi(0))+\gamma\|w\|_{\mathcal{L}_2})^2$. If $\exists T$ s.t. $\forall t>T$, $(\xi(t),\tau(t))\in\mathcal{X}\setminus\underline{\mathcal{A}}$, then $|\mathcal{T}_s|=2I_s$ for some finite $I_s\in\mathbb{N}$. Since $\forall t\in\mathbb{R}_0^+,\,W(t)\geq 0$, and $W(t_{2I_s}^s)=0$: $-\int_{t_{2I_s}^s}^{\infty}\frac{d}{dt}W(t)dt\leq 0$, and thus $\int_{t_{2I_s}^s}^{\infty}z_{\underline{\mathcal{A}}}^{\mathrm{T}}(t)z_{\underline{\mathcal{A}}}(t)dt\leq \gamma^2\int_{t_{2I_s}^s}^{\infty}w^{\mathrm{T}}(t)w(t)dt$. Therefore, it holds that $\|z_{\mathcal{A}}\|_{\mathcal{L}_2}^2\leq\|z_{\underline{\mathcal{A}}}\|_{\mathcal{L}_2}^2= \sum_{i=0}^{I_s-1}\int_{t_{2i}^s}^{t_{2i+1}^s}z_{\underline{\mathcal{A}}}^{\mathrm{T}}(t)z_{\underline{\mathcal{A}}}(t)dt +\int_{t_{2I_s}^s}^{\infty}z_{\underline{\mathcal{A}}}^{\mathrm{T}}(t)z_{\underline{\mathcal{A}}}(t)dt
\allowbreak+ \sum_{i=0}^{I_s-1}\int_{t_{2i+1}^s}^{t_{2i+2}^s}\allowbreak z_{\underline{\mathcal{A}}}^{\mathrm{T}}(t)z_{\underline{\mathcal{A}}}(t)dt <(\delta(\xi(0))+\gamma\|w\|_{\mathcal{L}_2})^2$. If $\exists T$ s.t. $\forall t>T$, $(\xi(t),\tau(t))\in\underline{\mathcal{A}}$, then $|\mathcal{T}_s|=2I_s+1$ for some finite $I_s\in\mathbb{N}$, and thus $\int_{t_{2I_s+1}^s}^{\infty}z_{\underline{\mathcal{A}}}^{\mathrm{T}}(t)z_{\underline{\mathcal{A}}}(t)dt=0$. Therefore, it holds that $\|z_{\mathcal{A}}\|_{\mathcal{L}_2}^2\leq\|z_{\underline{\mathcal{A}}}\|_{\mathcal{L}_2}^2= \sum_{i=0}^{I_s-1}\int_{t_{2i+1}^s}^{t_{2i+2}^s}z_{\underline{\mathcal{A}}}^{\mathrm{T}}(t)z_{\underline{\mathcal{A}}}(t)dt+ \int_{t_{2I_s+1}^s}^{\infty} z_{\underline{\mathcal{A}}}^{\mathrm{T}}(t)z_{\underline{\mathcal{A}}}(t) dt\allowbreak+\sum_{i=0}^{I_s}\int_{t_{2i}^s}^{t_{2i+1}^s}z_{\underline{\mathcal{A}}}^{\mathrm{T}}(t)z_{\underline{\mathcal{A}}}(t)dt <(\delta(\xi(0))+\gamma\|w\|_{\mathcal{L}_2})^2$.\qed \\
\textbf{Proof of Proposition \ref{lemma:mxupperbound}}
Following the proof of Theorem \ref{theorem:stabilityandperformance}, one has $\forall (\xi(t),\tau(t))\in C_H\setminus\underline{\mathcal{A}}$: $\frac{d}{dt}W(\xi(t),\tau(t))<-2\rho W(\xi(t),\tau(t))+w^{\mathrm{T}}(t)w(t)$. Apply the Comparison Lemma on the interval $[t_{2i}^s,T]$, where $T\in[t_{2i}^s,t_{2i+1}^s]$ to obtain $W(T)< W(t_0^s)+\frac{\|w\|_{\mathcal{L}_{\infty}}^2}{2\rho}$. When $(\xi(t),\tau(t))\in\underline{\mathcal{A}}$, $W(t)$ is bounded by $W(t)=0\leq 0.5\rho^{-1}\|w\|_{\mathcal{L}_{\infty}}^2$
 , and thus $W(t)\leq W(0)+\frac{1}{2\rho}\|w\|_{\mathcal{L}_{\infty}}^2,\,\forall(\xi(t),\tau(t))\in\mathcal{X}$. From the definition of $W(x,r)$ in (\ref{eq:newlpf}), together with the fact that $V(t)\geq\underline\lambda|\xi(t)|^2$, one obtains $\forall t\in\mathbb{R}_0^{+},\,
|\xi(t)|^2\leq\frac{W(0)+\frac{1}{2\rho} \|w\|_{\mathcal{L}_{\infty}}^2+\bar\lambda\bar\varrho^2\eta_{\min}^2}{\underline{\lambda}}$. Thus $m^i(t_k) \leq\eta_i^{-0.5}(t_k)(|\hat{u}^i(t_{k-1})|+|u^i(t_k)|) \leq\eta_i^{-0.5}(t_k)(|\xi(t_{k-1})|+|[C\,D]||\xi(t_k)|)$. Combining these bounds, it is clear that
(\ref{Eq:mxupperbound}) holds.\qed \\
\textbf{Proof of Proposition \ref{lemma:mmuupperbound}} Proof of Proposition \ref{lemma:mmuupperbound} is analogous to that of Proposition \ref{lemma:mxupperbound}.\qed

\bibliographystyle{plain}        
\bibliography{mybib}           

\end{document}